\documentstyle [floats,preprint,aps,eqsecnum]{revtex}

\global\arraycolsep=2pt         

\makeatletter
\def\footnotesize{\@setsize\footnotesize{9.5pt}\xpt\@xpt
\abovedisplayskip 10pt plus2pt minus 5pt
\belowdisplayskip \abovedisplayskip
\abovedisplayshortskip \z@ plus 3pt
\belowdisplayshortskip 6pt plus 2pt minus 2pt
\def\@listi{\topsep 6pt plus 2pt minus 2pt
\parsep 3pt plus 2pt minus 1pt \itemsep \parsep}}
\makeatother

\begin{document}

\title{Nuclear Electron Capture in a Plasma}

\author{Lowell S. Brown}

\address{
Department of Physics, University of Washington
\\
Seattle, Washington 98195
\\}

\author{R. F. Sawyer}

\address{
Department of Physics, 
University of California at Santa Barbara
\\
Santa Barbara, California 93106
\\}

\maketitle

\begin{abstract}

  We consider the electron density at the position of an ion of charge
  $Ze$ in a plasma under conditions approximating those in the core of
  the sun.  Numerical calculations have shown that the plasma effects
  on the density, over and above the ordinary Coulomb factors that are
  obtained in the absence of electron-electron interactions, are well
  represented by a reduction factor, $ \exp\{- Z e^2 \beta \kappa_D
  \}$, where $\beta$ is the inverse temperature and $\kappa_D$ is the
  Debye wave length. Although this factor is the direct analogue of
  the Salpeter enhancement factor for the fusion rates in stars, the
  elementary considerations that establish it in the fusion case are
  not applicable to the determination of the electron density and the
  resulting electron capture rates. We show analytically, through a
  sum rule that leads to a well-defined perturbative approach, that in
  the limit of Boltzmann statistics the Salpeter factor indeed
  provides the leading correction. We estimate residual effects, both
  from Fermi statistics and from short range terms.

\end{abstract}

\newpage

\section{Introduction and Summary}

The purpose of this paper is to provide a clear and rigorous
treatment of the process of nuclear electron capture in a plasma within
the mean field approximation. 

It is illuminating to begin by reviewing the effects of the
surrounding plasma on the fusion of positively charged ions. The
experience in looking at these plasma effects under solar (or weak
screening) conditions is that the correction calculated by Salpeter
(1954) is considerably larger than any other effect. The essence of
this correction is replacement of the Coulomb potential by the Debye
screened potential. Since the classical turning point for the ionic
barrier penetration is at a much smaller radius than the Debye radius,
it suffices to evaluate a screening energy correction to the barrier
penetration problem, an energy determined by taking the difference
between the unscreened and screened potentials as $r$ approaches zero.
This energy difference is given by $\Delta E = - e^2 Z_1 Z_2 \kappa_D
\ $ where $ Z_1$ and $Z_2$ are the charges of the two nuclei. Then the
obvious statistical argument gives the enhancement factor $ \exp\{
\beta e^2 Z_1 Z_2 \kappa_D \}$.

However it is not at all clear that this factor applies, as a matter
of principle, to the plasma corrections to the electron density at a
nucleus.  One difference is that we now deal with an attraction rather
than a repulsion, so that at least the language of the above
qualitative description must be changed. But the important difference
is that the thermal wavelength for the electron is much greater than
that for the ion; for the solar problem it is only a little smaller
than the Debye length.  The WKB approximation gives a theoretical
justification for considering the outer regions of the electronic
cloud surrounding the nucleus to be governed by a classical
statistical distribution determined by the temperature, chemical
potential, and local electrostatic potential.  
But for an electron at
a distance from the nucleus of one wavelength, the WKB approach is
not applicable. Thus the simplest argument for the Salpeter factor (now a
suppression) seems not to apply to the electron case.

Nevertheless, in the case of greatest current interest in solar
processes, electron capture in  $^7Be$, more detailed calculations of
electronic wave functions have given screening related reductions to the
$r=0$ electron density that
are only slightly less than would have resulted from a Salpeter formula. In
these calculations, beginning with those of Iben, Kalata and Schwarz
(1967), then sharpened by Bahcall and Moeller (1969), and by Johnson,
Kolbe, Koonin, and Langanke (1992),  the screening induced reduction
comes about in an way that we describe below. 

1) The Saha equation is used to determine the degree of occupation of
the bound states in $^7Be$ in the medium assuming pure Coulomb
electron--ion forces; for example, there turns out to be roughly a 20\%
probability of occupation for both the $m = \pm 1/2$
 1S states (Iben, Kalata, and Schwartz 1967). 
Leaving out screening, the 1S states give a  contribution
of about 35\% of the continuum contribution to the $r=0$ electron
density at the nucleus. The higher bound states in this picture have
appreciable occupation as well, and contribute another 6\% of the
continuum value.

2) When Debye screening is introduced, as shown by Bahcall and
Moeller (1967), the contribution of the continuum states to the $r=0$
density is changed by a very small amount, of the order of 1\%. 

3) The screening reduces the occupation factor for the 1S state by a bit;
it changes the wave function at $r=0$ of this state by quite a bit more,
with the end result that in the screened problem the 1S contribution is
reduced to about 20\% of the continuum contribution. As pointed out by
Gruzinov and Bahcall (1997), combining this reduction with the
complete removal of the higher bound states gives a net reduction due to
screening that is quite close to that predicted by a Salpeter formula. 

Gruzinov and Bahcall  address the problem through a quantum
diffusion equation elucidated by Feynman (1990) and applicable in the
case of Boltzmann statistics. The approach provides both a qualitative
understanding of why the Salpeter factor is a good approximation in the
high temperature limit and a computational framework for incorporating
the effects of classical plasma fluctuations. Unlike the other approaches
described above, it does not need to calculate occupation probabilities
from the Saha equation. (The Saha equation, and indeed the
``percentage occupancy'' that it describes, are not  well defined once
electrons interact with themselves.)

In the present work, which we view as complementary to that of
Gruzinov and Bahcall, we show analytically that in the Boltzmann limit
the Salpeter correction is indeed the leading correction in a well
defined perturbative approach. Furthermore, in the leading order of
approximation, Fermi statistics can be maintained at little cost,
since the generalization of the Salpeter multiplicative factor, in
going from pure Coulomb to screened, is a simple displacement of the
electron chemical potential $\mu_e$. Thus the Salpeter factor is
regained in the Maxwell-Boltzmann limit in which everything is
proportional to $\exp\{ \beta\mu_e\} $.
 
In establishing these results we do not need to separate bound
from continuum parts, or use a Saha equation. We
have also calculated leading small corrections to this basic result.
In contrast to the fluctuation corrections of Gruzinov and Bahcall
(1997), the additional terms depend on the quantum mechanics of the
plasma.

We turn now to describe our results; the mathematical details that
support them are relegated to the Appendix. 

As in the previous work, we exploit the large nucleus -- electron mass
ratio so that the nucleus may be treated as a fixed point of charge
$Ze$, with the capture rate proportional to the electron density at
this point which we take to be the coordinate origin.  In many-body
field theory language, this density is given by the thermal ensemble
average
\begin{equation}
D(\beta,\mu_e) = \langle \psi^\dagger(0) \psi(0) \rangle_\beta \,.
\end{equation}
Here\footnote{We employ energy units for the temperature $T$ and 
  use natural units in which Planck's constant $\hbar =
  1$.} $\beta = 1/ T$ is the inverse temperature and $\mu_e$ is the
chemical potential of the electrons.  This general formula includes all
possible corrections resulting from the electron-plasma interactions
to the electron density at the nucleus. The thermal expectation value
of the field operators is the coincident point limit of the two-point,
single-particle electron Green's function, including all plasma
interactions as well as having an additional fixed point charge $Ze$
at the coordinate origin. This single-particle Green's function has a
standard representation in terms of the single-particle irreducible
electron self-energy function $\Sigma$. The self-energy function may
be divided into two parts: 1. A part that contains all the terms
depicted by all the graphs which end with a single Coulomb photon line
connected to the electron line. These graphs define an effective,
external, local screened potential $V_S(r)$ in which the electron
propagates. In the small $Z$, dilute plasma limit, this potential is
the Coulomb potential produced by the heavy nucleus as modified by the
plasma polarization accounted for by the familiar ring graph sum. In
general, however, $V_S(r)$ contain all possible interactions of the
heavy nucleus with the plasma and all possible plasma interactions,
except that all these interactions are communicated in the end to the
electron by a single Coulomb photon exchange. 2. A part that contains
all other plasma effects. These entail at least two Coulomb photon
lines attached to the electron line. The leading correction in this
part is the exchange energy correction that is familiar in the
Hartree-Fock description of atoms. 

In this paper, we shall investigate only the corrections to the
electron density at the nucleus resulting from the screened
potential\footnote{Although we have given a precise definition of
  $V_S(r)$ in terms of the single-photon line reducible contribution
  to the electron self-energy function $\Sigma$, the following
  considerations apply to any local mean-field potential $V_M(r)$.
  However, any terms that are added to our definition of $V_S(r)$ must
  then be subtracted from the remainder of $\Sigma$, and the net
  effect of the resulting $\Sigma$ must be shown to be small.}
$V_S(r)$.  Thus the dynamics of the electron field operator $\psi({\bf
  r},\tau)$ is governed by the screened potential $V_S(r)$ with no
other particle-particle interactions. The electron field operator
satisfies the simple Schr\"odinger equation for a particle in the
potential $V_S(r)$. The Green's functions of the theory are of the
same form as those in the completely non-interacting theory except
that in their spectral representation, free-particle wave functions
are replaced by the corresponding Schr\"odinger eigenfunctions in the
potential $V_S(r)$.  Hence, with no further approximation, the density
reads
\begin{equation}
D_S(\beta,\mu_e) = \int dE \, n(E,\mu_e) \, \left| \psi_S(E;0) \right|^2 \,,
\label{ds}
\end{equation}
where the integration implies in addition a summation over possible
discrete bound states, and where 
\begin{equation}
n(E,\mu_e) = { 2 \over e^{ \beta ( E - \mu_e) } + 1 } \,,
\label{weight}
\end{equation}
with the 2 in the numerator accounting for the 2 spin states; 
$\psi_S(E;0)$ is the properly normalized Schr\"odinger wave function
for energy $E$. It should be emphasized that the result (\ref{ds})
automatically accounts for the proper weighting of the bound state
contributions, no additional ``Saha-like'' reasoning need be done.

In the limit in which the screened potential is
replaced by the Coulomb potential
\begin{equation}
V_C(r) = - { Z e^2 \over r} \,,
\end{equation}
we have the continuum wave functions for $ 0 \le E < \infty $ with
\begin{equation}
\left| \psi_C(E;0) \right|^2 = { Z m \over \pi a_0 } { 1 \over 1 - e^{
    - 2 \pi  \eta} } \,,
\end{equation}
in which $m$ is the electron mass, $E = p^2 / 2m$, $ a_0 = 1 / e^2 m$
is the electron Bohr radius, and $\eta = Z / a_0 p $ is the usual
Coulomb parameter. In addition, there is the infinite set of bound
state wave functions giving
\begin{equation}
\left| \psi_n(0) \right|^2 = { Z^3 \over \pi a_0^3 n^3 } \,,
\end{equation}
with the bound state energies
\begin{equation}
E_n = - { Z^2 e^2 \over 2 a_0 n^2 } \,.
\end{equation}
We shall denote the the electron density at the nucleus in this
Coulomb limit by $D_C(\beta,\mu_e)$, with
\begin{equation}
D_C(\beta,\mu_e) = \sum_{n=1}^\infty \left| \psi_n(0) \right|^2
n(E_n,\mu_e) + \int_0^\infty dE \, n(E,\mu_e) \,
\left| \psi_C(E;0) \right|^2 \,.
\label{coulb}
\end{equation}

We shall express the corrections in terms of the electron density
$\langle n_e \rangle_\beta $ in the plasma far away from the capturing
nucleus. Neglecting interacting plasma effects, this density is given
by
\begin{eqnarray}
\langle n_e \rangle_\beta &=& \int { (d^3 {\bf p}) \over (2\pi)^3 } 
n(E,\mu_e)
\nonumber\\
&=& 2 \lambda_e^{-3} e^{\beta\mu_e} \left[ 1 - { 1 \over 2\sqrt 2}
  e^{\beta\mu_e} + { 1 \over 3 \sqrt 3} e^{2\beta\mu_e} + \cdots \right] \,,
\label{eden}
\end{eqnarray}
where, in the second line $\lambda_e$ is the electron thermal
wavelength defined by
\begin{equation}
\lambda_e = \sqrt{ 2\pi\beta \over m} \,,
\label{le}
\end{equation}
and we have expanded the denominator in the Fermi-Dirac distribution
$n(E,\mu_e)$, performed the momentum integrals, and kept the first
two corrections to the classical statistics limit. For our numerical
corrections, we shall use the parameters stated by Gruzinov and Bahcall
(1997) which describe the solar interior at a distance 6\% away
from the sun's center. We shall also write the parameters in
essentially atomic units, except that we shall display the units. Thus
we take
\begin{equation}
\beta = 0.0215 \, \left( a_0 / e^2 \right) \,,
\label{temp}
\end{equation}
corresponding to a temperature $T = (e^2 / a_0) / 0.0215 = 1.27$
KeV. This temperature gives
\begin{equation}
\lambda_e = 0.368 \, a_0 \,.
\end{equation}
The electron density is taken to be
\begin{equation}
\langle n_e \rangle_\beta = 9.10 \, / a_0^3 \,,
\end{equation}
which gives by Eq.~(\ref{eden})
\begin{equation}
e^{\beta\mu_e} = 0.245 \,.
\end{equation}
The resulting Debye wave number defined\footnote{Accounting for
  Fermi-Dirac statistics, the electron contribution to the Debye wave
  number is given by $ \kappa_{D,e}^2 = 4\pi e^2 \partial \langle n_e
  \rangle_\beta / \partial \mu_e$. This correct definition reduces the
  total Debye wave number by 2\%, but it entails only a negligible
  1/2\% effect for $^7Be$ capture.} by $\kappa^2_D = 4\pi e^2 \beta 
\, 2 \langle n_e \rangle_\beta$ reads
\begin{equation}
\kappa_D = 2.22 \, / a_0 \,, \qquad\qquad \kappa_D^{-1} = 0.451 \, a_0
\,.
\end{equation}

We shall first demonstrate that the electron density
$D_D(\beta,\mu_e)$ at the nucleus for the Debye potential
\begin{equation}
V_D(r) = - { Z e^2 \over r} e^{ - \kappa_D r}
\label{dpot}
\end{equation}
may be expressed, to a good approximation, in terms of the
Coulomb limit and then describe the correction that arise from a more
accurate treatment of the screened potential. This relationship between
the Debye and Coulomb density follows from a sum rule (\ref{ssrule}) 
proved in the
Appendix\footnote{Note that the convergence of the
  integration for $ E \to \infty$ is delicate: The two squared wave
  functions must be subtracted at the same energy; their separate
  integrals do not exist.},
\begin{equation}
\int dE \, \left\{ \left| \psi_D(E;0) \right|^2 -
\left| \psi_C(E';0) \right|^2  \right\} = 
 { Z \kappa_D^2 \over 16 \pi a_0 } \,,
\label{srule}
\end{equation}
where
\begin{equation}
E' = E - Z e^2  \kappa_D \,.
\label{eprime}
\end{equation}
We make use of this sum rule to write
\begin{equation}
D_D(\beta,\mu_e) = \int dE \, n(E,\mu_e) \, \left| \psi_C(E';0) \right|^2 
+ R_D(\beta,\mu_e) \,,
\label{result}
\end{equation}
in which
\begin{equation}
R_D(\beta,\mu_e) = \int dE \, [n(E,\mu_e) - n(0,\mu_e)] 
\left\{ \left| \psi_D(E;0) \right|^2 - \left| \psi_C(E';0) \right|^2 
\right\} + { Z \kappa_D^2 \over 16 \pi a_0 } n(0,\mu_e) 
\label{rem}
\end{equation}
will be shown to be a quite small correction.  Translating the energy
in the integration in Eq.~(\ref{result}) to remove the displacement
shown in the definition (\ref{eprime}) of $E'$ and noting that this
translation just shifts the value of the chemical potential $\mu_e$ in
the weight (\ref{weight}), we see that Eq.~(\ref{result}) may be
expressed as
\begin{equation}
D_D(\beta,\mu_e) = D_C(\beta,\mu_e - Ze^2 \kappa_D)
+ R_D(\beta,\mu_e) \,.
\label{major}
\end{equation}

To keep simple analytic forms, we shall express the corrections in
terms of the bulk electron density $\langle n_e \rangle_\beta$ in the 
plasma. Including the first non-classical correction,
\begin{equation}
{ Z \kappa_D^2 \over 16 \pi a_0 } n(0,\mu_e) = 
\langle n_e \rangle_\beta
{ Z \kappa_D^2 \lambda_e^3 \over 16 \pi a_0 } \left[ 1 - 
  \left( 1 - { 1 \over 2 \sqrt 2} \right) e^{\beta\mu_e}  \right] \,.
\label{summ}
\end{equation}
Since the sum rule implies that on the average the difference between
$\left| \psi_S(E;0) \right|^2 $ and $ \left| \psi_C(E';0) \right|^2 $
is small, the integral in $R_D(\beta,\mu_e)$, 
\begin{equation}
I_D(\beta\mu_e) = \int dE \, 
\left\{ \left| \psi_D(E;0) \right|^2 - \left| \psi_C(E';0) \right|^2 
\right\} \left[ n(E,\mu_e) - n(0,\mu_e) \right] \,,
\label{int}
\end{equation}
should give a small contribution. Moreover, the sum rule has been used
to subtract $n(0,\mu_e)$ from $n(E,\mu_e)$, and this difference
becomes large only at energies that are large on the atomic scale,
large energies that are on the order of the temperature $T$. At large
energies, the terms in the perturbative development of the wave
functions converges rapidly and so this high-energy contribution can
be computed analytically. Said another way, this contribution gives
the leading term for small $\beta$ and, in view of Eq.~(\ref{temp}),
$\beta$ is indeed small in the relevant atomic units. For moderate
energies, the integration in Eq.~(\ref{int}) gives terms that are
linear in $\beta$. The high-energy contribution obtained in
Eq.(\ref{est}) of the Appendix is of order $\sqrt\beta$ which is
larger for small $\beta$. Since the Fermi-Dirac corrections to the
Maxwell-Boltzmann limit are small, we include only the first correction
in this high-energy contribution. Combining the result of
Eq.~(\ref{est}) with Eq.~(\ref{summ}), we obtain the leading
correction to the remainder
\begin{eqnarray}
  R_D(\beta,\mu_e) &=& \langle n_e \rangle_\beta {Z \kappa_D
    \lambda_e^3 \over 2 \pi a_0^2 } \Bigg\{ { a_0 \kappa_D \over 8}
  \left[ 1 - \left( 1 - { 1 \over 2 \sqrt 2} \right) e^{\beta\mu_e}
  \right] 
\nonumber\\ 
&& \qquad \qquad
+  \left( {3 Z \over 4} - { a_0 \kappa_D \over 12 } \right) 
\sqrt{ { 1 \over \pi} { \beta \kappa_D^2 \over 2m} } 
\left[ 1 - { 3 \sqrt{2} \over 4 }
    e^{\beta\mu_e} \right] \Bigg\} \,.
\end{eqnarray}
For the parameters listed above,
\begin{equation}
R_D(\beta,\mu_e) = 0.018 \, Z \{ 0.28 \, [ 1 - 0.16]  
+ 0.098 \, ( Z - 0.25 ) [ 1 - 0.26] \}
\langle n_e \rangle_\beta \,.
\label{correct}
\end{equation}
For $Z=4$, this gives $ R_D(\beta,\mu_e) = 0.04 \,
\langle n_e \rangle_\beta $, while as we shall see shortly, the
corresponding capture density $R(\beta\mu_e)$ is about $ 4 \, \langle
n_e \rangle_\beta$. Thus this leading additional correction to the
basic correction provided by the Salpeter factor is only 1\%, and any
further corrections should be smaller yet.

The Debye potential (\ref{dpot}) that we have been using has the
linear term $ - Ze^2 \kappa_D \,r $ in its short distance limit. As
the work in the Appendix shows, it is this term that gives the
non-vanishing value to the right-hand side of the sum rule
(\ref{srule}). No linear term in $r$ appears in the correct screened
potential $V_S(r)$ since such a term would give rise to an unphysical
screening charge density [ $ - \nabla^2 r = - 1/r$]. However, the
corrections that remove this linear term at extremely short distances
arise from wave numbers in the Fourier transform of the potential,
$\tilde V_S(k)$, that are of order $ 1 / \lambda_s$, where $\lambda_s$
is the thermal wave number of a species $s$ particle in the
plasma. For the ions in the plasma, the distance $\lambda_s$ is very
small in comparison with the other relevant distances in our problem,
and this cut off in not important. Indeed, the work of the Appendix
shows that this effect for the ions in the plasma is of order $ m /
M_s $ relative to the small corrections that we have already
displayed. But for the electrons in the plasma, the cut off at
$\lambda_e$ is as important as the other corrections that we have
displayed. This effect of the electrons in the plasma is computed in
the Appendix in the dilute plasma limit and to leading order in the
small parameter $ \beta Z^2 e^2 / 2 a_0$. In these limits, the
correction given by Eq.~(\ref{done}) reads
\begin{equation}
\Delta D(\beta,\mu_e) =  \langle n_e \rangle { Z \kappa_D^2 \lambda_e^3
  \over 8 \pi^3 a_0 } \, I \,,
\end{equation}
where $I$ is a parameter integral that has the numerical value 
$ I = 1.105$. For the parameter values used before, this gives
a negligible effect:
\begin{equation}
\Delta D(\beta,\mu_e) = 0.001 \, Z \, \langle n_e \rangle_\beta  \,.
\end{equation}

We have now shown that to within an accuracy on the order of 1\%,
the screened electron density $D_S(\beta,\mu_e)$ is given by the
simple Coulomb density, but with a translated chemical potential,
$D_C(\beta,\mu_e - Z e^2 \kappa_D)$. It remains to evaluate this main
term using Eq.~(\ref{coulb}). Following Gruzinov and Bahcall, we
define an enhancement factor which is the ratio of the electron
density at the nucleus calculated in various schemes to the average
electron density given in Eq.~(\ref{eden}), 
\begin{equation}
w = { D(\beta,\mu_e) \over \langle n_e \rangle_\beta } \,.
\end{equation}
We denote by $w_S$ the complete result of the shifted chemical
potential Coulomb density including the effect of Fermi-Dirac
statistics and including the small correction shown in
Eq.~(\ref{correct}). 
The same quantity, but using Maxwell-Boltzmann statistics, is labeled
$w_{S,B}$. The ratio with no screening corrections, the simple Coulomb
result is written as $w_C$ for the case of the full Fermi-Dirac
statistics and $w_{C,B}$ in the Maxwell-Boltzmann limit. Finally, the
results of Gruzinov and Bahcall (1997) will be denoted as $w_{GB}$. As
discussed previously, we use the parameters given by Gruzinov and
Bahcall in order to compare our results with theirs. The ratios given
by the various schemes as well as a direct comparison with their
results are displayed in the following table:

\bigskip
\begin{center}
\tabcolsep=1em
\begin{tabular}{ccccccc}
\hline\hline
Z       &    1    &    2    &    3    &    4    &    5    &    6 
\\ \hline
$w_S$   &  1.38   &  1.89   &  2.63   &  3.67   &  5.12   &   7.19
\\
$w_{S,B}$& 1.39   &  1.94   &  2.72   &  3.84   &  5.46   &   7.82
\\
$w_C$   & 1.43    &  2.05   &  2.96   &  4.29   &  6.22   &   9.04
\\
$w_{C,B}$& 1.45   &  2.11   &  3.10   &  4.59   &  6.84   &  10.3
\\
$w_{GB}$&  1.39   &  1.94   &  2.73   &  3.85  &   5.50   &  7.90 
\\ \hline
${w_{S,B} \over w_{GB}}$ &
           1.00   &  1.00   &  1.00   &  1.00   & 0.99    &   0.99
\\
${w_S \over w_{S,B}}$ &
           0.99   &  0.97   &  0.97   &  0.96   &  0.94   &   0.92
\\ \hline\hline
\end{tabular}
\end{center}
\bigskip

The ratio of ratios ${w_{S,B} / w_{C,B} }$ is essentially the Salpeter
factor $\exp\{- \beta Z e^2 \kappa_D \}$ which varies from 0.95 for
$Z=1$ through 0.83 for $Z=4$ to 0.75 for $Z=6$.  The small term $
R_D(\beta,\mu_e) $ displayed in Eq.~(\ref{correct}) gives a correction
to $w_{S,B}$ that varies from 0.4 \% for $Z=1$ through 1 \% for both
$Z=4$ and $Z=6$. The next-to-last row in the table, the ratio of
ratios ${w_{S,B} / w_{GB}}$ should be precisely unity if our
approximations were without error and the computation of Gruzinov and
Bahcall were precise.\footnote{We have recalculated Gruzinov and
  Bahcall's mean field values and found small discrepancies for $Z=1$
  and $Z=6$.  The corrected results are given in our table.} The last
row in the table, the ratio of ratios ${w_S / w_{S,B}}$, displays the
effect of Fermi statistics on the capture rate.

We conclude that the numbers from the calculations cited earlier are
sufficiently accurate to determine the electron capture in rate in
$^7Be$ to the precision needed for analysis of future solar neutrino
results.  The development presented in the present work first provides a
rigorous basic formulation that unambiguously describes the electron
density at the nucleus in the screened field approximation with no
need of any considerations of the Saha type. It then gives an
analytical way of understanding the main features of previous
calculations, as well as an approach that may be useful in addressing
related problems in the future.

\begin{center}

ACKNOWLEDGMENTS

\end{center}

We wish to thank the organizers of the Institute of Nuclear Theory 
``Solar Nuclear Fusion Rates'' Workshop (Seattle, 1997) for enhancing
our interest in electron capture. 
One of the authors (LSB) has had fruitful discussions on this topic
with L.~G. Yaffe. This work was supported, in part, by the
U.S. Department of Energy under Grant No. DE-FG03-96ER40956.

\newpage

\appendix

\section{Sum Rule, High-Energy Behavior}

Here we shall derive the sum rule and other results
used in the text. This will be done by examining the high-energy
behavior of S-wave, radial Green's functions defined by the
inhomogeneous differential equation
\begin{equation}
\left[ - {1 \over 2m} {d^2 \over dr^2} + V(r) - E \right]
{\cal G}(E; r,r') = \delta(r - r') 
\label{geq}
\end{equation}
together with outgoing wave boundary conditions. In particular, we
shall first compare the Green's function ${\cal G}_D(E; r,r')$ for the
Debye potential
\begin{equation}
V_D(r) = - { Ze^2  \over r} e^{- \kappa_D r} \,,
\label{dpott}
\end{equation}
with the Green's function ${\cal G}_C(E'; r,r')$ at the
translated energy
\begin{equation}
E' = { {p'}^2 \over 2m} = E - Z e^2   \kappa_D \,,
\end{equation}
for the Coulomb potential
\begin{equation}
V_C(r) = - {Ze^2 \over r} \,.
\end{equation}
This comparison is most easily done by noting that
\begin{equation}
{\cal G}_D(E; r,r') = {\cal G}_C(E'; r,r') - \int_0^\infty d  \bar r 
\, {\cal G}_C(E'; r, \bar r) \Delta V(\bar r) {\cal G}_D(E; \bar r,r') \,,
\label{ieq}
\end{equation}
where
\begin{eqnarray}
\Delta V(r) &=& V_D(r) - V_C(r) + E' - E
\nonumber\\
&=&  - Z e^2 \left[ {1 \over r} \left( e^{- \kappa_D r} - 1 \right) 
+ \kappa_D \right] \,.
\end{eqnarray}
This integral equation defines a Green's function ${\cal G}_D(E;
r,r')$ which obeys the proper differential equation (\ref{geq}) with
the Debye potential $V_D(r)$, and this Green's function is defined
with outgoing wave boundary conditions if these boundary conditions
are obeyed by the Coulomb Green's function ${\cal G}_C(E'; r,r')$.

The high-energy behavior of the Green's function may be obtained by
iterating the integral equation (\ref{ieq}) to form the perturbative
series
\begin{equation}
{\cal G}_D(E; r,r') = {\cal G}_C(E'; r,r') - \int_0^\infty d \bar r \,
{\cal G}_C(E'; r, \bar r) \Delta V(\bar r) {\cal G}_C(E'; \bar r,r')
+ \cdots \,.
\label{perts}
\end{equation}
The high-energy limit probes the short-distance limit of the
perturbation, $\Delta V(\bar r) \to - Z e^2 \kappa_D^2 \, \bar r / 2
$, and the high-energy limit of the Coulomb Green's function which (by
simple dimensional reasons) has an overall factor of $m/p$. The
Fourier transform involved in the $\bar r$-integration of the leading
$\bar r$ term in $\Delta V(\bar r)$ gives rise, in the high-energy
limit to another factor of $ 1 / p^2 $. Thus the successive terms in
the perturbative development are smaller by a factor of $ 1 / p^3 $ in
the high-energy limit, and, as we shall see, it suffices for our
purposes to retain only the first correction as shown in
Eq.~(\ref{perts}).

The Coulomb Green's function may be constructed in terms of solutions
to the S-wave Coulomb radial Schr\"odinger equation, the homogeneous
counterpart of the Green's function differential equation (\ref{geq}).
These solutions are confluent hypergeometric functions which have
standard integral representations. The functions we need may be
defined by
\begin{equation}
A(E;r) = -2ipr \int_0^\infty dt \, e^{ipr (2t+1) } t^{-i\eta} ( 1
+t)^{i\eta} \,,
\label{arep}
\end{equation}
and
\begin{equation}
B(E;r) = { r \over \Gamma(1 - i\eta) \Gamma(1 + i\eta) } \int_0^1 du
\, e^{ -ipr(2u-1) } u^{-i\eta} ( 1 - u)^{i\eta} \,,
\end{equation}
where
\begin{equation}
\eta = {Z e^2 m \over p } = { Z \over p a_0 } \,.
\end{equation} 
It is a straight forward matter to check that these integral
representations obey the Coulomb S-wave radial Schr\"odinger
equation\footnote{The application of this differential equations
  results in integrals of total derivatives that vanish.}. It is also 
not difficult to establish the limits
\begin{equation}
r \to 0 \,: \qquad\qquad A(E;r) \to 1 \,,
\end{equation}
\begin{equation}
r \to 0 \,: \qquad\qquad B(E;r) \to r \,,
\end{equation}
and the asymptotic behavior
\begin{equation}
r \to \infty \,: \qquad A(E;r) \to e^{ipr} \left( { i \over 2pr}
\right)^{-i\eta} \Gamma(1 -i\eta)  \,.
\end{equation}
Thus $B(E;r)$ is the regular solution and $A(E;r)$ has outgoing
waves. Moreover, the $r \to 0$ limit and the constancy of the
Wronskian give
\begin{equation}
W = {d A(E;r) \over dr} B(E;r) - A(E;r) { d B(E;r) \over dr} = -1 \,.
\end{equation}
Accordingly, the Coulomb Green's function has the construction
\begin{equation}
{\cal G}_C(E; r,r') = 2m \, A(E;r_>) \, B(E;r_<) \,,
\label{cgreen}
\end{equation}
where $ r_> \,,\, r_<$ are the greater and lessor of $r \,,\, r'$.

The results that we need are obtained by examining the high-energy
behavior of
\begin{equation}
\Delta(E) = \lim_{r,r' \to 0} { 1 \over r} \left[ {\cal G}_D(E; r,r') -
  {\cal G}_C(E'; r,r') \right] { 1 \over r'} \,.
\end{equation}
To the order of accuracy that we need, Eq.~(\ref{perts}) and the
Coulomb Green's function construction (\ref{cgreen}) give 
\begin{equation}
\Delta(E) = - (2m)^2 \int_0^\infty dr \, \Delta V(r) A(E;r)^2 \,.
\label{pertt}
\end{equation}
In the high-energy limit, the Coulomb parameter $ \eta = Z / pa_0$
becomes small, and so the integral representation (\ref{arep}) gives
\begin{equation}
A(E';r)^2 \approx e^{2ip'r} - 4 p' r \, \eta' 
\int_0^\infty dt \, e^{2ip'r (1 +  t)} \ln\left( t \over 1+t \right) \,.
\end{equation}
Moreover, since the high-energy limit involves only the short-distance
behavior of the perturbing potential, we may approximate
\begin{equation}
\Delta V(r) \approx - Z e^2 \left[ {1\over 2} \kappa_D^2 \, r - 
{1 \over 6} \kappa_D^3 \, r^2 \right] \,.
\end{equation}
To our order, only the first term here contributes in the $\eta'$
correction to $A(E';r)^2$, and we have
\begin{equation}
\Delta(E) = Z e^2 (2m)^2 \left\{ { \kappa_D^2 \over 2 (2ip')^2 } + {
    \kappa_D^3 \over 3 (2ip')^3 }
+ {4 \kappa_D^2 p' \eta' \over (2ip')^3 } \int_0^\infty 
{dt \over (1+t)^3 } \ln \left( { t \over 1+t } \right) \right\} \,.
\end{equation}
The integral that appears here is reduced to an elementary integral by
the variable change $ 1+t = 1/x$. Thus, after a little algebra, we
find that the leading high-energy limit is given by
\begin{equation}
E \to \infty \,: \qquad \Delta(E) = - { Z \kappa_D^2 \over 4 a_0 E } 
+ i { Z \over a_0^3 E^{3/2} } A\,,
\label{limit}
\end{equation}
in which
\begin{equation}
A =\left[  { a_0^2 \kappa_D^2 \over 12} - { 3 Z a_0 \kappa_D \over 4}
\right]  \sqrt{ \kappa_D^2 \over 2m}
\label{aresult}
\end{equation}

To obtain the desired sum rule, we note that
the Green's functions have a spectral representation. Since the radial
functions that we have used are related to the total wave function by
\begin{equation}
\psi(E; r) = { u(E;r) \over r} \, Y^0_0 = { u(E;r) \over r} { 1 \over
  \sqrt{4\pi} } \,,
\end{equation}
this spectral representation yields
\begin{equation}
{ \Delta(E) \over 4\pi  } = \int d \bar E \, { |\psi_S(\bar E;0)|^2 -
    |\psi_C(\bar E';0)|^2 \over \bar E - E - i \epsilon} \,,
\end{equation}
where the integration implicitly includes a sum over all bound
states. In view of the high-energy limit (\ref{limit}),  we conclude that
\begin{equation}
\int d E \, \left\{ |\psi_S( E;0)|^2 -
    |\psi_C( E';0)|^2 \right\} = + { Z \kappa_D^2 \over 16 \pi a_0 } \,,
\label{ssrule}
\end{equation}
which is the sum rule used in the text. Since
\begin{equation}
{\rm Im} \, { 1 \over \bar E - E - i \epsilon} = 
\pi  \delta ( \bar E - E) \,,
\end{equation}
we also have
\begin{equation}
|\psi_S( E;0)|^2 - |\psi_C( E';0)|^2 = {{\rm Im} \, 
\Delta(E) \over 4 \pi^2} \,,
\label{im}
\end{equation}
which we shall now use to obtain the corrections presented in the text.

The result (\ref{im}) expresses the integral in the remainder 
$R_D(\beta,\mu_e)$ defined in Eq.~(\ref{int}) as
\begin{equation}
I_D(\beta,\mu_e) = \int dE \, { {\rm Im} \Delta(E) \over 4 \pi^2 }
\Big[ n(E, \mu_e) - n(0,\mu_e) \Big] \,.
\end{equation}
Since, in the relevant atomic units, the inverse temperature $\beta$
is quite small, we shall evaluate this integral in the small $\beta$
limit, which should give the leading correction. The integration over
finite energies gives a result that is first order in $\beta$, but, as
we shall soon see, the integration region of large $\beta$ gives a
larger contribution for small $\beta$, a contribution of order
$\sqrt\beta$. This larger contribution is given by the high-energy
limit (\ref{limit}). Since the departure from classical statistics is
small, we shall retain only the first-order correction in
$\exp\{\beta\mu_e\}$ and thus compute
\begin{eqnarray}
I_D(\beta\mu_e) &\approx& { Z A \over 4 \pi^2 a_0^3 } \, 2 e^{\beta\mu_e}
\int_0^\infty { d E \over E^{3/2} } \left[ \left( e^{-\beta E} - 1
  \right) - e^{\beta\mu_e} \left( e^{-2\beta E} - 1 \right) \right]
\nonumber\\
&=& - { Z A \over 4 \pi^2 a_0^3 } 2 e^{\beta\mu_e} 2 \sqrt{\beta\pi}
\left[ 1 - \sqrt{2} e^{\beta\mu_e} \right] \,.
\end{eqnarray}
Using Eq.~(\ref{eden}) to express this in terms of the bulk electron
density gives
\begin{equation}
  I_D(\beta\mu_e) \approx - \langle n_e \rangle_\beta  \, { Z A
    \lambda_e^3 \over 2 \pi a_0^3 }  \sqrt{\beta \over \pi} \left[ 1 - { 3
      \sqrt{2} \over 4 } e^{\beta\mu_e} \right] \,,
\end{equation}
with the result (\ref{aresult}) finally yielding
\begin{equation}
I_D(\beta\mu_e) \approx \langle n_e \rangle_\beta 
{ Z \kappa_D \lambda_e^3 \over 2 \pi a_0^2 } 
\left[ 1 - { 3 \sqrt{2} \over 4 } e^{\beta\mu_e} \right] 
\left( { 3 Z \over 4}  - { a_0 \kappa_D \over 12 }  \right) \sqrt{ { 1 \over
      \pi} { \beta \kappa_D^2 \over 2m} } \,.
\label{est}
\end{equation}

We turn at last to examine the correction to the electron density at
the nucleus that comes from the difference between the Debye potential
$V_D(r)$ and a more accurate screened potential $V_S(r)$. Since the
two potentials have the same long-distance behavior, their difference
becomes important only at high energies where the effects of the
Coulomb interaction become small.  Hence, a good estimate is obtained
by replacing the Coulomb function $A(E';r)$ by the plane wave
$\exp\{ipr\}$ in Eq.~(\ref{pertt}):
\begin{equation}
\bar \Delta_1(E) = - (2m)^2 \int_0^\infty dr \, e^{2ipr} 
\left\{ V_S(r) - V_D(r) \right\} \,.
\end{equation}
The Fourier transform of the screened potential may be written in the
general form\begin{equation}
\tilde V_S(k) = - Z e^2 { 4 \pi \over k^2 + 4 \pi \Pi(k) } \,.
\end{equation}
With $ 4\pi \Pi(k) = \kappa_D^2$, this gives the Debye potential. Hence
\begin{eqnarray}
V_S(r) - V_D(r) &=& Z e^2 \int 
{ (d {\bf k}) \over (2\pi)^3}  e^{ i {\bf k} \cdot {\bf r} }
{ 4\pi \over k^2 + \kappa_D^2 } { 4\pi \left[ \Pi(k) - \Pi(0) \right]
 \over k^2 + 4\pi \Pi(k) }
\nonumber\\
&\approx& 8 Z e^2 \int_0^\infty {dk \over k^2} \, {\sin kr \over kr} 
\left[ \Pi(k) - \Pi(0) \right] \,,
\end{eqnarray}
where is the second approximate equality we have performed the angular
integration and neglected the denominator corrections that are of
order $\kappa_D^2$ and do not contribute to the leading high-energy
behavior. Using
\begin{equation}
\int_0^\infty {dr \over r} \, e^{2ipr} \, \sin kr =
- { i \over 2} \ln \left[ { 2p -k \over 2p + k } \right] \,,
\end{equation}
and thus obtain
\begin{equation}
{\rm Im} \, \bar \Delta_1(E) = { 16 m Z \over a_0 } \int_0^\infty { dk
  \over k^3 } \left[ \Pi(k) - \Pi(0) \right] \ln \left| { 2p -k \over
    2p + k } \right| \,.
\end{equation}

Again for simplicity using Maxwell-Boltzmann statistics, this gives the
correction to the electron density at the nucleus of
\begin{equation}
\Delta D(\beta,\mu_e) = 2 e^{\beta\mu_e} \int_0^\infty { pdp \over m} 
\exp\left\{ - \beta { p^2 \over 2m } \right\}  
{ {\rm Im} \, \bar \Delta_1(E) \over 4 \pi^2 } \,.
\end{equation}
We now represent the plasma polarization function $\Pi(k)$ by its
one-loop, ring graph approximation. Using the results of Brown and
Sawyer, their Eq's.~(A41) and (A42), we find that
\begin{equation}
\Pi(k) - \Pi(0) =  {\sum}_s e^2_s \beta \langle n_s
\rangle_\beta \int_0^1 dt \, 
\left[ \exp\left\{ - { \beta k^2 \over 2M_s} t ( 1 - t ) \right\} 
- 1 \right] \,,
\end{equation}
where the sum runs over all species $s$ of the particles in the
plasma. We perform the integration over $p$ by writing $ k = 2p x$ to obtain
\begin{eqnarray}
\Delta D(\beta,\mu_e) &=& - 2 e^{\beta\mu_e} 
{\sum}_s e^2_s \beta \langle n_s \rangle_\beta { Z \over 2\pi^2 a_0 }
\int_0^\infty { dx \over x^3 } \int_0^1 dt
\nonumber\\
&& \qquad\qquad \ln \left| { 1 -x \over 1 + x } \right| 
\ln \left[ 1 + 4 x^2 t (1-t) { m \over M_s} \right] \,.
\label{last}
\end{eqnarray}

The ratio $ m / M_s$ is very small for the ions in the plasma. In the
ionic case, the final logarithm above gives only a term of order $ m /
M_s$. Thus only the electrons in the plasma give a significant
contribution to this high-energy correction. Since the electron
density is half the total number density in the completely ionized
plasma, we may write
\begin{equation}
4 \pi e^2 \beta \langle n_e \rangle_\beta = { 1 \over 2} \kappa^2_D
\,.
\end{equation}
The $t$ integration in Eq.~(\ref{last}) is elementary, and we secure
at last
\begin{equation}
\Delta D(\beta,\mu_e) =  \langle n_e \rangle { Z \kappa_D^2 \lambda_e^3
  \over 8 \pi^3 a_0 } \, I \,,
\label{done}
\end{equation}
in which $I$ is the analytically intractable integral
\begin{equation}
I = - \int_0^\infty { dx \over x^4} 
\left\{ \sqrt{1 + x^2 } \ln \left[ \sqrt{ 1 + x^2} + x \right] 
- x \right\} 
\ln \left| { 1 - x \over 1 + x } \right| \,,
\end{equation}
whose numerical integration gives $ I = 1.105$ .

\newpage

\begin{center}
REFERENCES
\end{center}

Bahcall, J.\ N., \& Moeller, C.\ P.\ 1969, ApJ 155, 511

Brown, L.\ S., \& Sawyer, R.\ F.\ 1997, Rev.\ Mod.\ 
Phys.\ 69, 411 (astro-ph/9610256)

Feynman, R.\ P.\ 1990, Statistical Mechanics (Reading:
Addison-Wesley), Chapter 2, especially p. 48

Gruzinov, A.\ V., \& Bahcall, J.\ N., to be published (astro-ph/9702065)

Iben, I., Jr., Kalata, K., \& Schwartz, J.\ 1967, ApJ 150, 1001

Johnson, C.\ W., Kolbe, E., Koonin, S.\ E., \& Langanke, K.\ 1992, ApJ
392, 320

Salpeter, E.\ E. 1954, Australian J. Phys., 7, 373

\end{document}